\documentclass[aps,prb,twocolumn,floats,showpacs]{revtex4}
\usepackage{epsfig}
\usepackage{amsmath}
\usepackage{amssymb}
\begin{document}
\title{Ultraviolet Probing of Quantum Crossbars}
\author{ I. Kuzmenko$^{1}$, S. Gredeskul$^{1}$, K. Kikoin$^{1}$, and
Y. Avishai$^{1,2}$\\ }
\affiliation {$^1$Department of Physics and $^2$Ilse Katz Center,
Ben-Gurion University, Beer-Sheva, Israel }
\date{\today}

\begin{abstract}
Ultraviolet (UV) scattering on quantum crossbars (QCB) is an
effective tool for probing QCB spectral properties, leading to
excitation of QCB plasmon(s).  Experimentally, such a process
corresponds to sharp peaks in the frequency dependence of the
differential scattering cross section.  The peak frequency
strongly depends on the direction of the scattered light.  As a
result, $1D\to 2D$ crossover can be observed in the scattering
spectrum. It manifests itself as a splitting of single lines into
multiplets (mostly doublets). The splitting magnitude increases
with interaction in QCB crosses, while the peak amplitudes
decrease with electron-electron interaction within a QCB
constituent.
\end{abstract}
 \maketitle

%%%%%%%%%%%%%%%%%%%%%%%%%%%%%%%%%%%%%%%%%%%%
\section{Introduction}\label{sec:Intro}
%%%%%%%%%%%%%%%%%%%%%%%%%%%%%%%%%%%%%%%%%%%%
Quantum crossbars (QCB) is a novel artificial nano-object which
represents actually a double $2D$ grid formed by two superimposed
crossing arrays of parallel conducting quantum wires
\cite{Avron,Avishai,Guinea1,Melosh}, molecular chains\cite{Luo} or
metallic single wall carbon nanotubes (SWCNT)
\cite{Melosh,Rueckes,Mukho1,Dalton}.  Similar structures with the same
crossbar geometry also arise naturally as, e.g., crossed striped
phases of doped transition metal oxides \cite{Tranq}.  The QCB
mechanical flexibility together with the possibility of exciting some
of its constituents (nanotube, a single wire) by external electric
field and the existence of bistable conformations of others (molecular
chain\cite{Tseng}) makes QCB one of the most attractive architectures
for designing molecular-electronic circuits for computational
application \cite{Luo,Rueckes,Heath}.  One more attractive property of
QCB, its optical activity in a wide frequency region from infrared
(IR) to ultraviolet (UV) one, will be discussed in the present
paper.\\

From a geometrical point of view, QCB is, in a sense, an object with
intermediate dimensionality.  The two dimensional character of QCB as
a whole, together with the one dimensional character of its
constituents enables studying the crossover from a $1D$ Luttinger
liquid (LL) behavior to a $2D$ Fermi liquid (FL) behavior.  Indeed, a
single wire or nanotube possesses the LL-like spectrum
\cite{Bockrath,Egger}.  An array of parallel wires is still a LL-like
system (qualified as a sliding phase \cite{Mukho1}) provided the pure
electrostatic interaction between adjacent wires is taken into
account.  However, inter-wire tunneling turns the electronic spectrum
of an array to that of $2D$ FL\cite{Wen1,Schultz1}.  A system of two
crossing arrays (QCB) manifests the same behavior.  Being coupled only
by capacitive interaction in the crosses, they have similar
low-energy, long-wave properties characterized as a crossed sliding LL
phase\cite{Mukho1,KGKA1}.  Inter-array electron tunneling destroys the
LL behavior at low energies and results in dimensional crossover to
$2D$ FL\cite{Mukho1,Guinea2,Mukho2}.  Here the control parameter which
rules the crossover is the tunneling intensity.\\

Outside the low energy region, the intermediate character of the QCB
dimensionality leads to nontrivial spectral properties which cannot be
treated in terms of purely $1D$ or purely $2D$ electron liquid theory. 
QCB with only electrostatic interaction in the crosses possesses the
LL zero energy fixed point.  Far from this point, the system conserves
a Bose character in the absence of interwire tunneling but the
dimensionality of its Bose excitations (plasmon modes) becomes
intermediate between $1D$ and $2D$ in spite of the fact that they live
in the $2D$ Brillouin zone (BZ)\cite{KGKA1,KGKA2}.  These QCB plasmons
can be treated as a set of dipoles within the QCB constituents.  In a
single wire, the density of their dipole momenta is proportional to
the LL boson field $\theta(x)$ ($x$ is the coordinate along the wire). 
In QCB, there are two sets of coupled dipoles.  They form a unique
system which manifests either $1D$ or $2D$ properties depending on
details of the relevant experiment.  This results in a new dimensional
crossover.  It differs from the LL to FL transition mentioned above and is
ruled by other control parameters (quasimomentum, energy, frequncy of
an external field and so on).  Some possibilities of observing such
crossover in transport measurements (which give information about the
nearest vicinity of the LL fixed point at $(q,\omega,T)\to 0$) were
discussed in Ref.\onlinecite{Mukho1}.  Other crossover effects such as
appearance of non-zero transverse space correlation functions and
periodic energy transfer between arrays ("Rabi oscillations") were
studied in Ref.\onlinecite{KGKA2}.  Observation of these effects
probes the QCB spectral properties outside the LL fixed point (i.e.
well beyond the crossing sliding phase region).\\

A rather pronounced manifestation of thess kinds of dimensional
crossover is related to QCB response to an external ac electromagnetic
field.  The two main parameters characterizing the plasmon spectrum in
QCB are the velocity $v$ of plasmons in a single wire and the QCB
period $a$ (we assume that periods in the two basic directions are
equal).  These parameters define both the typical plasmon wave numbers
$Q_0=2\pi/a$ and their typical frequencies $\omega_0= vQ_0$.  Choosing
$v\approx 8\cdot 10^{7}$~cm/sec and $a\approx 20$~nm, according to
Refs.[\onlinecite{Egger,Rueckes}], one finds that plasmon frequencies
lie in the far infrared (IR) region $\omega\sim 10^{14}$~sec$^{-1}$,
while characteristic order of plasmon wave vectors is $Q_0\sim
10^{6}cm^{-1}.$\\

A very natural method to study the QCB spectrum is by means of optical
spectroscopy.  The characteristic values of QCB frequencies and wave
vectors determine two possible directions of such an experimental
observation.  The first one is IR spectroscopy of QCB where
\textit{the frequency} of an external ac field lies at the same region
as the QCB frequency.  The second one is an UV scattering on QCB where
\textit{the wave vector} of a scattered field lies in the same region
as that of the QCB wave vectors.\\

The effectiveness of IR spectroscopy of QCB were studied in Refs. 
\onlinecite{KGKA3,K}.  Here, the IR light wave vector $k$ is three
orders of magnitude smaller than the characteristic plasmon wave
vector $Q_0$ (the plasmon velocity $v$ is much smaller than the light
velocity $c$).  Therefore, an infrared radiation incident on an {\em
isolated} array, can excite plasmon only with $\omega=0,$ or simply
speaking, cannot excite plasmons at all.  However in the QCB geometry,
each array serves as a diffraction lattice for its partner, giving
rise to Umklapp processes with Umklapp vectors say $(nQ_0,0)$ with an
integer $n.$ As a result, excitation of plasmons in the BZ center with
frequencies $\omega=nvQ_0$ is possible.\\

To excite QCB plasmons with non zero wave vectors, an additional
diffraction lattice with period $A>a$ coplanar with the QCB can be
used.  Here the diffraction field contains space harmonics with wave
vectors $2\pi M/A$ ($M$ integer), that enables one to eliminate the
wave vector mismatch and to scan plasmon spectrum within the BZ. In
the general case, one can observe single absorption lines forming two
equidistant sequences.  However, in case where the wave vector of the
diffraction field is oriented along some resonance directions,
additional absorption lines appear.  As a result, an equidistant
sequences of split doublets can be observed in the main resonance
direction (BZ diagonal).  This is the central concept of dimensional
crossover mentioned above with direction serving as a control
parameter.  In higher resonance directions, absorption lines form an
alternating series of singlets and split doublets demonstrating new
type of dimensional crossover.  The latter occurs in a given direction
with frequency as a control parameter.\\

One more version of IR QCB spectroscopy is related to the study of QCB
placed onto a semiconductor substrate\cite{KGKA3,K}.  It occurs that a
capacitive contact between the QCB and the substrate does not destroy
the LL character of the long wave QCB excitations.  However, the
dielectric losses of a substrate are drastically modified due to
diffraction processes on the QCB superlattice.  QCB - substrate
interaction results in the appearance of additional Landau damping
regions of the substrate plasmons.  Their existence, form, and density
of losses are strongly sensitive to variation of the QCB lattice
constant.\\

The IR based methods mentioned above are not very convenient from two
points of view.  First, as it was mentioned, one needs an additional
diffraction lattice to tune the light wave vector and that of the QCB
plasmon.  Second, they probe QCB spectrum only in some discrete
points.  The alternative method of studying QCB spectrum by {\em
ultraviolet} (UV) {\em light scattering} is the subject of the present
paper.  The advantages of this method are evident.  It does not
require any additional diffraction lattice.  It probes QCB spectrum in
a continuous region of wave vectors.  Finally, its selection rules
differ from those for IR absorption.  This gives rise to the
observation of additional spectral lines not visible in IR
experiments.\\

In this paper we formulate the principles of UV spectroscopy for QCB
and study the main characteristics of scattering spectra.  The paper
is organized as follows.  In subsection \ref{sec:spectr}, we briefly
describe double square QCB and quantum numbers associated with its
excitations.  In the next subsection, \ref{sec:Light}, we discuss
light scattering on QCB and and present basic equations describing
this process.  The main results of the paper are contained in Section
\ref{sec:Scat} where we classify the basic types of the scattering
indicatrices (angular diagrams of differential cross section)
corresponding to various detector orientations.  The results obtained
are summarized in the Conclusions.  Technical details are concentrated
in two Appendices.  Representation of QCB plasmons in terms of array
plasmons is explained in Appendix \ref{sec:Kinem}.  Appendix
\ref{sec:Inter} is devoted to the derivation of an effective QCB-light
interaction.  The basic formulas for differential cross section of
light scattering are concentrated in this Appendix.\\
%%%%%%%%%%%%%%%%%%%%%%%%%%%%%%%%%%%%%%%%%%%%
\section{Light scattering on QCB}\label{sec:QCB}
%%%%%%%%%%%%%%%%%%%%%%%%%%%%%%%%%%%%%%%%%%%%%%%%%%%%%%%%%%%%%
%%%%%%%%%%%%%%%%%%%%%%%%%%%%%%%%%%%%%%%%%%%%
\subsection{QCB: Geometry, Hamiltonian, Spectrum}\label{sec:spectr}
\begin{figure}[htb]
\begin{center}
\includegraphics[width=65mm,height=80mm,angle=0,]{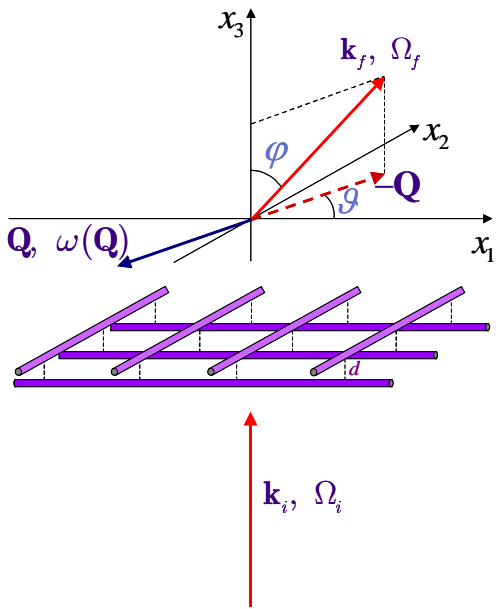}
\caption{The scattering process geometry (all notations are
explained in the text).} \label{X-Scatt}
\end{center}
\end{figure}

%%%%%%%%%%%%%%%%%%%%%%%%%%%%%%%%
A square QCB is formed by two periodically crossed perpendicular
arrays of $1D$ quantum wires or carbon nanotubes.  The two arrays are
lying on two parallel planes separated by an inter-plane distance $d$
(see Fig.  \ref{X-Scatt}).  They are labeled by indices $j=1,2$ and
wires within the first (second) array are labeled by an integer index
$n_{2}$ ($n_{1}$).  The coordinate system in Fig.  \ref{X-Scatt} is
chosen such that (i) the axes $x_{j}$ and the corresponding basic unit
vectors ${\bf e}_{j}$ are oriented along the $j$-th array; (ii) the
$x_{3}$ axis is perpendicular to the QCB plane; (iii) the $x_{3}$
coordinate is zero for the first array and $d$ for the second one.  A
single wire is characterized by its radius $r_{0}$, length $L$, and LL
interaction parameter $g$.  The minimal nanotube radius is
$r_{0}\approx 0.35$~nm\cite{Louie}, maximal nanotube length is
$L\approx 1$~mm, and the LL parameter is estimated as $g\approx 0.3$
\cite{Egger}.  Typical QCB period and inter-array distance are
$a\approx 20$~nm, $d\approx 2$~nm\cite{Rueckes} so that usually
$r_{0}\ll d\ll a\ll L.$\\

The QCB Hamiltonian is
\begin{equation}
    H_{QCB}=H_{1}+H_{2}+H_{12}.
    \label{HamiltTot}
\end{equation}
The first term ${H}_{1}$ describes LL in the first array
\begin{eqnarray}
{H}_{1} & = &
         \frac{\hbar v}{2}\sum_{{n}_{2}}
              \int\limits_{-L/2}^{L/2} {dx}_{1}
              \biggl\{
                   g{\pi}_{1}^{2} \left( x_1,
                   {n}_{2}a,0\right)\nonumber
                   \\
               &+ &    \frac{1}{g}
                   \left(
                        {\partial}_{{x}_{1}}
                        {\theta}_{1}
                        \left(
                             {x}_{1},{n}_{2}a,0
                        \right)
                   \right)^2
         \biggr\},
         \label{H_1}
\end{eqnarray}
where $\theta_{1}(x,y,z)$ and $\pi_{1}(x,y,z)$ are two the
conventional canonically conjugate boson fields of the first array. 
Its counterpart $H_{2}$ describes the second array and is obtained
from $H_{1}$ by permutation $1\leftrightarrow 2$ and a replacement
$0\rightarrow d$ in the arguments of the fields.\\
 %%%%%%%%%%%%%%%%%%%%%%%%%%%%%%%%%%%%%%%%%%%%%%%%%%%%%%%%%%%%%
%%%%%%%%%%%%%%%%%%%%%%%%%%%%%%%%%%%%%%%%%%%%%%%%%%%%%%%%%%%%%%

The third term in Eq.(\ref{HamiltTot}) $H_{12}$ describes the
capacitive inter-array interaction.  It has a separable form
\begin{eqnarray}
   H_{12}  &=&  V_{0}\sum\limits_{{n}_{1},{n}_{2}}
            \int _{-L/2}^{L/2}dx_1 dx_2\nonumber\\
            &\times&\zeta\left(
                 \frac{x_1-n_1a}{r_{0}}
                 \right)
            \zeta\left(
                 \frac{n_2a-x_2}{r_{0}}
                 \right)\nonumber \\
         &\times& \partial_{x_1}\theta_1(x_1,n_2a,-d)
         \partial_{x_2}\theta_2(n_1a,x_2,0),\nonumber\\
         &&V_{0}=\frac{2e^{2}}{d}.
     \label{Interaction}
\end{eqnarray}
and is localized around the QCB crosses\cite{KGKA2}. This form of
inter-array interaction is actually a limiting form of an exact
inter-array Coulomb interaction for $r_{0}/d \ll 1$ taking into
account electron screening in a single nanotube\cite{Sasaki}.\\

The Hamiltonian (\ref{HamiltTot}), (\ref{H_1}), (\ref{Interaction})
describes QCB in the absence of tunneling.  In real QCB with typical
parameters mentioned above, an interwire tunneling is rather small
\cite{Rueckes}.  Coulomb blockade accompamying charging of adjacent
tubes in a process of electron transition from one tube to another,
suppresses these processes.  Only electron-hole cotinneling without
recharging is possible, and this cotunneling slightly renormalizes the
coefficient $V_{0}$ in Eq.  (\ref{Interaction}). The corresponding
relative correction is of order of $10^{-4}.$ \\

\begin{figure}[htb]
\begin{center}
\includegraphics[width=75mm,height=60mm,angle=0,]{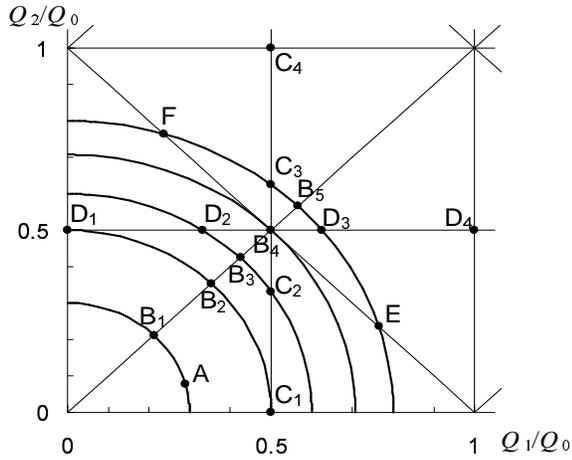}
\caption{Part of the reciprocal space.  The small square $Q_j/Q_0\leq
0.5$ is the quarter of the first QCB BZ. High symmetry lines are
parallel to the coordinate axes.  Resonant lines are parallel to BZ
diagonals.  The arcs coming over the point $B_{1},$ $B_{2},$ $B_{3},$
$B_{4},$ $B_{5},$ correspond to different wave numbers of the excited
plasmons: $|Q/Q_{0}|=0.3; \ \ 0.5; \ \ 0.6; \ \ 0.7; \ \
0.8.$}\label{BZNew}
\end{center}
\end{figure}

The coordinates $d$ and $0$ along the third spatial direction $x_{3}$
enter the QCB dynamics as parameters.  It is natural to expect that
QCB possesses of $2D$ spectral properties which can be classified
by $2D$ quasi-momentum.  The basic vectors of the $2D$
reciprocal superlattice are $Q_0{\bf e}_j,$ $j=1,2,$ so that any
vector of the reciprocal superlattice ${\bf m}$ is a sum ${\bf m}={\bf
m}_1+{\bf m}_2,$ where ${\bf m}_j=m_jQ_0{\bf e}_j$ with integer
${m}_j$.  The first BZ of QCB is $|Q_j|<Q_0/2$ (see Fig. 
\ref{BZNew}).  An arbitrary vector ${\bf Q}=(Q_1,Q_2)$ of reciprocal
space can be written as ${\bf Q}={\bf q}+{\bf m},$ where ${\bf
q}=(q_1,q_2)$ belongs to the first BZ. QCB eigenstates are classified
by quasi-momentum ${\bf q}$ and $2D$ band number $S$.\\

However, the specific QCB geometry makes its spectral properties
rather unusual.  Consider an {\em isolated} array 1.  Within the
($x_1,x_2$) plane, its excitations are described by a pair of $2D$
coordinates $(x_1,n_2a)$, i.e. a continuous longitudinal coordinate
$x_1$ parallel to the ${\bf e}_1$ direction, and its discrete transverse
partner $n_2a$ parallel to ${\bf e}_2.$ As a result, the longitudinal
component $Q_1=q_1+m_1Q_0$ of the excitation momentum changes on the
entire axis $-\infty < Q_1 < \infty$ while its transverse momentum
$q_2$ is restricted to the region $|q_2|<Q_0/2$.  Thus an eigenstate
(plasmon) of the first array is characterized by the vector ${\bf
Q}_1={\bf q}+{\bf m}_1=(Q_1,q_2)$ and the frequency
\begin{equation}
    \label{freq}
    \omega_{1}({\bf Q}_1)=v|Q_1|
    \nonumber
\end{equation}
which depends only on the longitudinal component $Q_1$ of the momentum
${\bf Q}_{1}$.  Similar description for the second array is obtained by
replacing $1\leftrightarrow2.$ \\

%%%%%%%%%%%%%%%%%%%%%%%%%%%%%%%%%%%%%%%%%%%%%%%%%%%%%%%%%%%%%
%%%%%%%%%%%%%%%%%%%%%%%%%%%%%%%%%%%%%%%%%%%%%%%%%%%%%%%%%%%%%%
In the case of "empty lattice" (QCB with infinitesimal interaction
between arrays, $V_{0}\to 0$) the complementary second array produces
infinitely small periodic potential which affects the first array. 
The vector ${\bf m}_1$ becomes a reciprocal $1D$ lattice vector and
${\bf q}$ becomes the $2D$ quasi-momentum.  The first array
excitations are labeled by this $2D$ quasi-momentum ${\bf q}$ and $1D$
band number
\begin{equation}
    s_1=\left[\frac{2|Q_1|}{Q_0}\right]+1
    \nonumber
    \label{bandnum}
\end{equation}
(square brackets denote an integer part of the number). The
corresponding dispersion law has the form

\begin{eqnarray}
 \omega_{s_1}({\bf q})\equiv&&\omega_1({\bf Q}_1)=v|Q_1|= \nonumber\\
   &&vQ_0\left(
    \left[\frac{s_1}{2}\right]
    +(-1)^{s_1-1}\frac{|q_1|}{Q_0}
    \right).
 \nonumber
 \label{omega}
\end{eqnarray}
As in the previous paragraph, by replacing $1\leftrightarrow2,$ we
obtain formulas describing eigenstates of the second array embedded
into an "empty lattice".  Thus an arbitrary vector ${\bf Q}={\bf
q}+{\bf m}_1+{\bf m}_2=(Q_1,Q_2)$ of the reciprocal space generates
two vectors ${\bf Q}_1={\bf q}+{\bf m}_1=(Q_1,q_2)$ and ${\bf
Q}_2={\bf q}+{\bf m}_2=(q_1,Q_2).$ The ``empty QCB'' eigenstates are
labelled by one of these vectors and its array number.  Each one of
the eigenstates propagates along the corresponding array.\\

The dimensionless strength of the inter-array interaction
(\ref{Interaction}) of QCB with typical values of the parameters
is small,
\begin{equation}\label{small}
\phi=\frac{gV_{0}r_{0}^{2}}{\hbar va}\approx 0.007.
\end{equation}
As a result, QCB eigenstates approximately conserve the same quantum numbers as the
(unperturbed) eigenstates of an ``empty QCB''.  These numbers are
array numbers $j=1,2,$ and the corresponding vectors ${\bf Q}_j.$ In
the general case, there are two QCB eigenstates associated with each wave
vector ${\bf Q}$ of reciprocal space.  In zeroth approximation,
these QCB eigenstates coincide with the ``empty QCB'' eigenstates
\begin{eqnarray}
    \label{one-boson}
    |1,{\bf Q}_1\rangle &= &a^{\dag}_{1,{\bf Q}_1}|0\rangle,\nonumber\\
    |2,{\bf Q}_2\rangle &= &a^{\dag}_{2,{\bf Q}_2}|0\rangle,
\end{eqnarray}
where $a^{\dag}$ is an array of plasmon creation operators and
$|0\rangle$ is the bosonic vacuum.  The corresponding eigenfrequencies
coincide with the unperturbed frequencies 
\begin{equation}
    \omega_{j}({\bf Q}_j)=v|Q_{j}|, \ \ j=1,2,
    \label{freqs}
\end{equation} 
in the same approximation.  In higher orders of perturbation, the
r.h.s. of Eq.  (\ref{one-boson}) includes additional terms like
$a^{\dag}_{1,{\bf Q}_1+{\bf m}}|0\rangle,$ where ${\bf m}$ is an
arbitrary reciprocal lattice vector.  However, the coefficients before
these terms are of higher order of magnitude with respect to the small
inter-array interaction (\ref{small}).\\

Such a simple description fails at the points ${\bf Q}$ lying on two
specific groups of lines in reciprocal space (see Appendix
\ref{sec:Kinem} for details).  Nevertheless, even at these points QCB
plasmons are represented as a finite superposition of array plasmons
with coefficients of order unity.  As the point ${\bf Q}$ moves away
from a specific line, only one of these coefficients survives while
all other decrease rapidly.  So more complicated structure of QCB
plasmons at the specific lines, presented in Appendix \ref{sec:Kinem},
turns to the standard description (\ref{one-boson}) within narrow
transition strips which adjoin to the specific lines.  Out of these
strips, i.e. in major part of reciprocal space (including the nearest
vicinities of the specific lines), a point ${\bf Q}$ completely
determines the structure of all possible QCB plasmons with momentum
${\bf Q}.$\\
%%%%%%%%%%%%%%%%%%%%%%%%%%%%%%%%%%%%%%%%%%%%%%%%%%%%%%%%%%%%%%%%%%%%
\subsection{Light scattering}\label{sec:Light}
\begin{figure}[htb]
\begin{center}
\includegraphics[width=70mm,height=55mm,angle=0,]{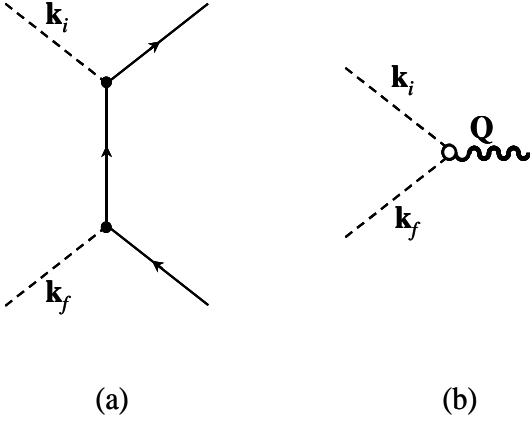}
 \caption{(a) Second order diagram describing light scattering on QCB.
 Solid lines correspond to fermions, whereas dashed lines are related
 to photons.  The vertices are described by the interaction
 Hamiltonian (\ref{H-int-initial}).  (b) Effective photon-plasmon
 interaction.  QCB excitation is denoted by wavy line.}
 \label{Feinmann-Diagramm}
\end{center}
\end{figure}

%%%%%%%%%%%%%%%%%%%%%%%%%%%%%%%%%%%%%%%%%%%%%%%%%%%%%%%%%%%%%
The simplest process contributing to Raman-like light scattering is an
annihilation of an incident photon and creation of a scattered photon
together with a QCB plasmon (Fig.  \ref{Feinmann-Diagramm}a).  In
terms of initial electrons, this is in fact a second order process. 
Since the energies of incident and scattered photons significantly
exceed the electron excitation energy in the nanotube, one may
consider the emission/absorption process as an instantaneous act (see
Appendix \ref{sec:Inter}).  This process may be treated as an 
inelastic photon scattering accompanied by emission or absorption of a
plasmon (Fig.  \ref{Feinmann-Diagramm}b).\\
\vspace{20mm}

%%%%%%%%%%%%%%%%%%%%%%%%%%%%%%%%%%%%%%%%%%%%%%%%%%%%%%%%%%%%%%%%%%%
%%%%%%%%%%%%%%%%%%%%%%%%%%%%%%%%%%%%%%%%%%%%%%%%%%%%%%%%%%%%%%%%%%%%
Let ${\bf k}_i$ (${\bf k}_f$) and $\Omega_i=ck_i\equiv c|{\bf
k}_{i}|,$ ($\Omega_f=ck_f$) be momentum and frequency of the incident
(scattered) photon.  For simplicity we restrict ourselves by the case
of normal incidence ${\bf k}_{i}=(0,0,k_{i}).$ Direction of the
scattered wave vector is characterized by a unit vector ${\bf
n}(\varphi,\vartheta),$ where $\varphi$ and $\vartheta$ are polar and
azimuthal angles in the spherical coordinate system with polar axis
directed along ${\bf e}_3$ direction.  The momentum of the excited QCB
plasmon is ${\bf Q}=-{\bf K},$ where ${\bf K}$ is the projection of the
scattered photon momentum ${\bf k}_{f}$ onto the QCB plane.  This process
is displayed in Fig.  \ref{X-Scatt}.  If the QCB plasmon frequency
$\omega({\bf Q})$ coincides with the photon frequency loss
$\omega=\Omega_i-\Omega_f,$ then a detector oriented in a scattered
direction ${\bf n}$ will register a sharp well pronounced peak at the
frequency loss $\omega=\omega({\bf Q})$.  The frequency loss is much
smaller than the incident and scattered photon frequencies.  Therefore
in what follows we use the same notation $k$ for both $k_{i}$ and
$k_{f}$ where it is possible.  Scanning the frequency loss $\omega,$
(or, equivalently, the modulus of the scattered wave vector $k_{f}$)
for a fixed detector orientation ${\bf n},$ two or more (up to six for
a square QCB) such peaks can be observed.  The number of peaks and
their location depend strongly on the azimuthal angle $\vartheta$ in
the QCB plane.  Scanning this angle, one can change the number of
observable peaks.  This is yet another manifestation of dimensional
crossover mentioned above.\\

The scattering differential cross section is smooth function of
$\omega.$ However, in zeroth approximation, only the peaks survive
(smooth background is very low because of weak inter-array
inetraction).  Therefore in a chosen approximation, the scattering
cross section is characterised by a number of peaks, their positions
and intensities.  Calculation of these quantities as functions of the
frequency loss $\omega$ and azimuthal angle $\vartheta$ for various
fixed values of the polar angle $\varphi,$ is the main goal of our
study.\\

 To perform quantitative analysis, we should derive an expression for
 the scattering cross section.  Details of all calculations are given
 in Appendix \ref{sec:Inter}.  Here we present only the main steps of
 the derivation.  We start from the Hamiltonian
\begin{equation}
 h_{nl} = \frac{ev_F}{c}
     \int\frac{dx_1d\gamma}{2\pi}
     \psi_{\alpha}^{\dag}
     {\bf{A}}\cdot
     {\boldsymbol\sigma}_{\alpha,\alpha'}
     \psi_{\alpha'},
       \label{H-int-initial}
\end{equation}
which describes interaction between a single nanotube oriented
e.g. along ${\bf e}_{1}$ (the first array nanotube) and an 
external electromagnetic field.  The field is described by its
vector potential in the Landau gauge
\begin{equation}
    {\bf{A}}=A_1{\bf{e}}_1+A_2{\bf{e}}_2+A_3{\bf{e}}_3.
    \label{VecPot}
\end{equation}€ 
The indices
$\alpha=a,b$ enumerate sublattices in a honeycomb carbon sheet,
$(r,\gamma)$ are polar coordinates in the $(x_{2},x_{3})$ plane,
$\psi_{\alpha}(x_{1},\gamma)\equiv\psi_{\alpha}(x_{1},r_{0},\gamma)$
and $\psi_{\alpha}^{\dag}$ are slowly varying electron field operators
at the nanotube surface $r=r_0,$ and the vector of Pauli matrices
${\boldsymbol\sigma}$ is
\begin{eqnarray}
\boldsymbol\sigma&=&
 {\bf{e}}_1\sigma_x
+{\bf{e}}_\gamma\sigma_y\nonumber\\
{\bf{e}}_{\gamma}&=&
-{\bf{e}}_2\sin\gamma
+{\bf{e}}_3\cos\gamma.\nonumber
\label{sigma-matrix}
\end{eqnarray}
The light wavelength is much larger than the nanotube radius, so the
vector potential can be taken at its axis, $ {\bf A}(x_1,0,0)$.  \\

Such form of a nanotube-light interaction leads to the following
expression for an effective QCB-light interaction Hamiltonian
\begin{eqnarray}
 H_{int} &=&
 \frac{\sqrt{2}}{4k}
  \frac{e^2}{\hbar c}
  \left(\frac{v_F}{c}\right)^2
 \sum_{n_2} \int dx_1
 \times\nonumber\\&&\times
 \partial_{x_1}\theta_1(x_1,n_2a)
 {\bf{A}}_1^2(x_1,n_2a,0)
 +\nonumber\\&+&
 \frac{\sqrt{2}}{4k}
  \frac{e^2}{\hbar c}
  \left(\frac{v_F}{c}\right)^2
 \sum_{n_1} \int dx_2
 \times\nonumber\\&&\times
 \partial_{x_2}\theta_2(n_1a,x_2)
 {\bf{A}}_2^2(n_1a,x_2,d).
 \label{Eff_Int_QCB}
\end{eqnarray}
Here
\begin{equation}
  {\bf{A}}_j({\bf r})={\bf{A}}({\bf r})+\left(\sqrt{2}-1
  \right)A_j({\bf r}){\bf e}_j, \ \ j=1,2,
  \nonumber
  \label{A-J}
\end{equation}
are two effective vector potentials affecting the two arrays and
$A_{j}({\bf r})$ are two (out of three) carthesian components of the
full vector potential (\ref{VecPot}).  The main object of our interest
is the scaled differential cross-section $\sigma(\omega,{\bf n})$ of
the scattering defined by
\begin{equation}
    {d\sigma}=L^{2}
    \frac{g}{4\pi ka}
    \left(\frac{e^{2}}{\hbar c}\right)^{2}
    \left(\frac{v_{F}}{c}\right)^{4}
    \sigma(\omega,{\bf n}){d\omega do},
    \label{eq:cross1}
\end{equation}
where $L^{2}$ is the QCB area and $do=\sin\varphi d\varphi d\vartheta.$
A standard procedure applied to the Hamiltonian (\ref{Eff_Int_QCB}),
leads to the following form of the scaled cross section
\begin{equation}
    \label{Fin_Cross}
    \sigma(\omega,{\bf n})=\frac{1}{4k}
    \sum_{P}
    \overline{|\langle P|\textrm{h}|0 \rangle_p|^2}
      \delta(\omega-\omega_P).
\end{equation}
Here $\omega=\Omega_i-\Omega_f$ is the frequency loss and
\begin{eqnarray}
    \label{Fin_Int}
    &&\textrm{h}=
    -\sum_j
    \frac{K_{j}}{\sqrt{|K_{j}|}}\times\nonumber\\
    &&P_{j;\lambda_f,\lambda_i}(\varphi,\vartheta)
    \left(a_{j,{\bf K}_{j}}+
    a^{\dag}_{j,-{\bf K}_{j}}
    \right)
\end{eqnarray}
is the interaction Hamiltonian reduced to the subspace of QCB states. 
Summation is performed over all one-plasmon states $|P\rangle.$ The
vector ${\bf K}_{j}$ is obtained from ${\bf K}_f$ in the same way as
the vector ${\bf Q}_j$ is obtained from ${\bf Q},$ and
$P_{j;\lambda_f,\lambda_i}(\varphi,\vartheta),$ $j=1,2,$ are
polarization matrices.  In the basis $(\|,\bot)$ they are
\begin{eqnarray}
    \label{Pol_Matr1}
    P_1(\varphi,\vartheta)&=&
    -\left(%
    \begin{array}{cc}
    2\sin\vartheta & i\cos\vartheta \\
    2i\cos\vartheta \cos\varphi & \sin\vartheta \cos\varphi .\\
    \end{array}
    \right),\nonumber\\
    P_2(\varphi,\vartheta)&=&P_1\Big(\frac{\pi}{2}+\vartheta,\varphi\Big).
\end{eqnarray}
Equations (\ref{Fin_Cross}) - (\ref{Pol_Matr1}) serve as a basis for the
subsequent analysis.\\
%%%%%%%%%%%%%%%%%%%%%%%%%%%%%%%%%%%%%%%%%%%%%%%%%%%%%%%%%%%%%%%%%%%%%%%
%%%%%%%%%%%%%%%%%%%%%%%%%%%%%%%%%%%%%%%%%%%%%%%%%%%%%%%%%%%%%
\section{Scattering cross section}\label{sec:Scat}
%%%%%%%%%%%%%%%%%%%%%%%%%%%%%%%%%%%%%%%%%%%%%%%%%%%%%%%%%%%%%
%%%%%%%%%%%%%%%%%%%%%%%%%%%%%%%%%%%%%%%%%%%%%%%%%%%%%%%%%%%%%
\subsection{Cross section: Basic types} \label{sec:Types}
%%%%%%%%%%%%%%%%%%%%%%%%%%%%%%%%%%%%%%%%%%%%%%%%%%%%%%%%%%%%%

According to Eqs.  (\ref{Fin_Cross}), (\ref{Fin_Int}), in order to
contribute to the cross-section (\ref{Fin_Cross}) in a fixed detector
orientation at ${\bf n},$ an excited QCB plasmon $|P\rangle $ must
contain at least one of two single-array states $|1,-{\bf
K}_{1}\rangle $ or $|2,-{\bf K}_{2}\rangle $.  Analysis shows that
there are five basic types of excited QCB plasmons depending on the
location of the point ${\bf K}.$ Here we describe these types of
plasmons and the corresponding structure of the differential cross
section of scattering.  In this description, beside the polar angle
$\varphi$ of the scattered photon wave vector ${\bf k}_{f},$ we use, in
a sense, a mixed representation.  It is based on the excited plasmon
momentum ${\bf Q}=-{\bf K}$ and the azimuthal angle $\vartheta$ of the
transverse component of the stattered photon wave vector ${\bf K}$.\\

%\vspace{40mm}
{\bf i.  General case: The point ${\bf Q}$ lies away from both the
high symmetry lines and the resonant lines.} This case is illustrated
by the point $A$ in Fig.  \ref{BZNew}.  There are two QCB plasmons
$|P\rangle,$ Eqs.  (\ref{one-boson}), with frequencies (\ref{freqs}),
which contribute to the scattering.  The differential cross-section is
a sum of two peaks centered at these frequencies.  After averaging
over initial and final polarizations, it has the form
\begin{eqnarray*}
    \sigma(\omega,{\bf n}) =
       F_{1}({\bf n})
       \delta(\omega-\omega_{1})+F_{2}({\bf n})
       \delta(\omega-\omega_{2}),
\end{eqnarray*}
where the functions
\begin{eqnarray*}
  F_1({\bf n})=&&F(\varphi,\vartheta)
  =
  |\sin\varphi\cos\vartheta|\times\\
  &&\Big[
       1-
       \frac{3}{4}\sin^2\varphi\cos^2\vartheta+
       \frac{1}{4}\cos^2\varphi
  \Big],\\
  F_2({\bf n})=&&F\Big(\varphi,\vartheta+\frac{\pi}{2}\Big),
\end{eqnarray*}
describe universal angle dependences of the peak amplitudes.  The
functions $F_{1,2}$ are related to the corresponding array plasmons. 
Each one of them vanishes when the scattered photon is perpendicular
to the corresponding array.  However these functions describe the
scaled cross section.  The absolute value amplitude of each peak has
an additional factor ${\displaystyle \frac{g}{4\pi
ka}\left(\frac{e^2}{\hbar
c}\right)^{2}€\left(\frac{v_F}{c}\right)^4}.$ Strong electron-electron
interaction in a nanotube corresponds to small values of the Luttinger
parameter $g$ and therefore suppresses the scattering cross section.\\

{\bf ii.  Inter-band resonance in one of the arrays: the point ${\bf
Q}$ lies on a high symmetry line of only one array.} This case is
illustrated by the points $C_{2,3}$ ($1$-st array) and $D_{2,3}$
($2$-d array) in Fig.  \ref{BZNew}.  Consider for example point $C_2$
where $Q_1=Q_0/2,\ \ Q_2\neq nQ_0/2.$ Here \emph{three} QCB plasmons
contribute to the scattering.  The first is a $|2,Q_2\rangle $
plasmon, (\ref{one-boson}), with frequency $\omega_2,$ Eq. 
(\ref{freqs}).  The other two are even or odd superpositions of the
$1$-st array states (Eq.  (\ref{states_1}) with $j=1$) of the two
first zones with eigenfrequencies (\ref{freq_b_1}).  Due to weakness
of the inter-array interaction, three peaks of the scattering cross
section form a singlet $\omega_2$ and doublet
$\omega_{1g},\omega_{1u}$.  After averaging over initial and final
polarizations, the cross section has the form
\begin{eqnarray*}
    \sigma(\omega,{\bf n}) &=&\frac{1}{2}
       F_{1}({\bf n})
       [\delta(\omega-\omega_{1g})+
       \delta(\omega-\omega_{1u}))]\\
       &+&F_{2}({\bf n})
       \delta(\omega-\omega_{2}).
\end{eqnarray*}
\\

{\bf iii.  Inter-band resonance in both arrays: The point ${\bf Q}$ is
a crossing point of two high symmetry lines away from all resonant
lines.} This case is illustrated by the points $C_4$ and $D_4$ in Fig. 
\ref{BZNew}.  Consider for example point $C_4$.  Here $Q_1=Q_0/2,\ \
Q_2= 2Q_0/2,$ and \emph{four} QCB plasmons contribute to the
scattering.  The first pair consists of even and odd superpositions of
the $1$-st array states of the first and the second bands.  These
states and their frequencies are described by Eqs. 
(\ref{high_cross}), (\ref{freq_b}) with $j=1.$ The second pair
consists of the same superpositions of the $2$-d array states from the
second and third bands and is described by the same equations with
$j=2.$ As a result, \emph{four} peaks, which form two doublets
(\ref{freq_b}), $j=1,2$, can be observed.  After averaging over
initial and final polarizations the cross section is
\begin{eqnarray*}
    \sigma(\omega,{\bf n})=
  \frac{1}{2}F_{1}({\bf n})
       [\delta(\omega-\omega_{1g})+
       \delta (\omega-\omega_{1u}))]\\+
  \frac{1}{2}F_{2}({\bf n})
       [\delta(\omega-\omega_{2g})+
       \delta\big(\omega-\omega_{2u}))].
\end{eqnarray*}
\\

{\bf iv.  Inter-array resonance: The point ${\bf{Q}}$ lies only on one
of the resonant lines away from the high symmetry lines.} This case is
illustrated by the points $B_{1-3,5},\ \ E,$ and $F$ in Fig. 
\ref{BZNew}.  Here the QCB plasmons which contribute to the scattering
are two even and two odd superpositions of the first and second array
states (\ref{first_1}) whose eigenfrequencies form two doublets
(\ref{freq_a_1}).  As in the previous case, the scattering cross
section contains \emph{four} peaks which form two doublets.  After
averaging over initial and final polarizations the cross section is
\begin{eqnarray*}
    \sigma(\omega,{\bf n}) =&&
  \frac{1}{2}
  F_{1}({\bf n})
       [\delta(\omega-\omega_{g}({\bf Q}_1))+
    \delta(\omega-\omega_{u}({\bf Q}_1))]+\nonumber\\
       &&\frac{1}{2}F_{2}({\bf n}) [\delta(\omega-\omega_{g}({\bf
       Q}_2))+\delta(\omega-\omega_{u}({\bf Q}_2))].
\end{eqnarray*}
\\

Thus, the inter-array splitting is proportional to the central small
parameter of the theory, $\phi$ (see Eq.  (\ref{small})).  For the set
of parameters described in the beginning of subsection
\ref{sec:spectr}, the interband splitting defined by Eq. 
(\ref{freq_b_1}), is five times smaller because it contains an
additional factor $\phi a/r_0.$\\

{\bf v.  Inter-array and inter-band resonance: The point ${\bf{Q}}$
lies at the intersection of two resonant lines.} There is only one
such point $B_4$ in Fig.  \ref{BZNew}.  In the general case, where the
parameter $n\neq 0$ for both crossing resonant lines (the point $B_4$
is {\em not} the case), the QCB plasmons involved in Raman scattering
form two quartets.  The first quartet consists of four symmetrized
combinations (\ref{quartet}) of the single-array states.  QCB
eigenstates for the second quartet are obtained from these equations
by replacing $1\leftrightarrow 2.$ The corresponding eigenfrequencies
are described by Eqs.  (\ref{evar}), (\ref{odar}).  The scattering
cross section in this case contains six peaks, and two of them are
two-fold degenerate
\begin{eqnarray*}
    \sigma(\omega,{\bf n})=
  \frac{1}{4}
  F_{1}({\bf n})[2\delta(\omega-\omega_{g,u/g}({\bf Q}_1))+\\
       \delta(\omega-\omega_{uu}({\bf Q}_1))+
       \delta(\omega-\omega_{ug}({\bf Q}_1)]+\\
       \frac{1}{4}
     F_{2}({\bf n})[2\delta(\omega-\omega_{g,u/g}({\bf Q}_2))+\\
          \delta(\omega-\omega_{uu}({\bf Q}_2))+
          \delta(\omega-\omega_{ug}({\bf Q}_2)].
\end{eqnarray*}
The point $B_4$ in Fig.  \ref{BZNew} lies on the main resonance line
with $n=0.$ Here $|Q_1|=|Q_2|,$ the frequencies of both quartets
coincide, and the scattering cross section contains one four-fold
degenerate peak and a symmetric pair of its two-fold degenerate
satellites.\\

This classification of all types of excited plasmons enables us
to describe completely the UV scattering on QCB.\\

%%%%%%%%%%%%%%%%%%%%%%%%%%%%%%%%%%%%%%%%%%%%%%%%%%%%%%%%%%%%%
\subsection{Scattering indicatrices}\label{sec:Rosettes}
%%%%%%%%%%%%%%%%%%%%%%%%%%%%%%%%%%%%%%%%%%%%%%%%%%%%%%%%%%%%%
In this subsection we describe the results of the scattering
process with the help of a family of scattering indicatrices.\\

To explain the indicatrix structure, we start with some preliminary
arguments.  Consider the case where the detector is tuned to the
frequency $\Omega_f$ and is oriented in the direction ${\bf n}.$ These
parameters uniquely determine a point ${\bf K}= -{\bf Q}$.  There are
two ways of scanning QCB plasmons.  The first way is to scan through
the polar angle $\vartheta.$ The corresponding points ${\bf K}$ in
Fig.  \ref{BZNew} form an arc with radius $K=(\Omega_f/c)\sin\varphi.$
The second way is to tune the detector frequency $\Omega_f.$ We are
interested in frequency loss of order of the plasmon frequencies
within the two - three lowest bands.  This loss is of order of
$\Omega_f(v/c)\ll\Omega_f.$ Therefore in this case the point ${\bf K}$
remains in its place (with a very good accuracy).\\

Each scattering indicatrix corresponds to a circular arc in Fig. 
\ref{BZNew} and the structure of this indicatrix is completely
determined by the arc radius $K.$ The indicatrix represents a set of
curves displayed in polar coordinates with polar angle, which
coincides with the azimuthal angle $\vartheta$ used above, and
(dimensionless) radius $\omega/(vQ_0),$ where $\omega$ is the
frequency loss.  Each point of the indicatrix corresponds to an
excitation of a QCB plasmon and therefore to a sharp peak in the
scattering cross section.  The number of peaks depends on the polar
angle.  Scanning the azimuthal angle $\vartheta$ results in changing
the number of peaks.  This is one more example of dimensional
crossover in QCB (see Ref.\onlinecite{KGKA3} for similar effects in IR
spectroscopy).\\

\begin{figure}[htb]
\begin{center}
\includegraphics[width=60mm,height=50mm,angle=0,]{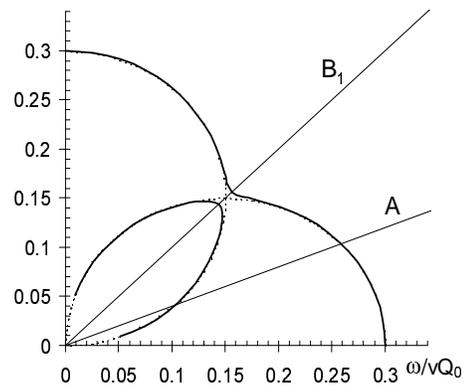}
\caption{Positions of the scattering peaks for $|K/Q_0|=0.3.$ The
doublet in the resonance direction (point $B_{1}$ in Fig. 
\ref{BZNew}) is well pronounced.}\label{Kll_03}
\end{center}
\end{figure}

We start with the case of smallest radius $k\sin\varphi=-0.3Q_{0}$
(arc $AB_1$ in Fig.  \ref{BZNew}).  Here all points beside the point
$B_{1}$ are points of general type ({\bf i}).  Each one of them, e.g.
point $A$, corresponds to excitation of two plasmons in the two arrays
and therefore leads to two separate peaks in the scattering spectrum
(see Fig.  \ref{Kll_03}).  The peaks corresponding to the point $A$ in
Fig.  \ref{BZNew} lie at the ray defined by the angle $\theta_{A}.$
The point $B_{1}$ corresponds to inter-array resonance ({\bf iv}) and
in this direction a split doublet can be observed.\\

%\vspace{40mm}
%%%%%%%%%%%%%%%%%%%%%%%%%%%%%%%%%%%%%%%%%%%%%%%%%%%%%%%%%%%%%%
\begin{figure}[htb]
\begin{center}
\includegraphics[width=60mm,height=50mm,angle=0,]{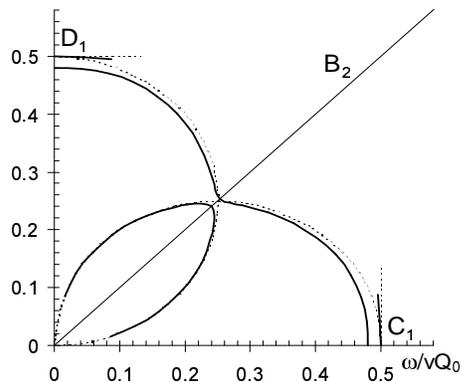}
\caption{Positions of the scattering peaks for $|K/Q_0|=0.5.$ Two
doublets appear at the BZ boundaries (points $C_{1}$ and $D_{1}$
in Fig. \ref{BZNew}).}\label{Kll_05}
\end{center}
\end{figure}

%%%%%%%%%%%%%%%%%%%%%%%%%%%%%%%%%%%%%%%%%%%%%%%%%%%%%%%%%%%%%
In the next case $k\sin\varphi=-0.5Q_{0}$ (arc $C_1B_2D_1$ in Fig. 
\ref{BZNew}), as in the previous one, in general directions one can
observe two single lines which form a split doublet in the resonant
direction $B_2$.  However at the final points $C_1$ and $D_1$ the arc
touches the high symmetry lines.  Here the low frequency single line
vanishes (there is no scattering at $Q=0$) while the high frequency
line transforms into a doublet because of an inter-band resonance
({\bf ii}) in one of the arrays (Fig.  \ref{Kll_05}).\\

%\vspace{40mm}
%%%%%%%%%%%%%%%%%%%%%%%%%%%%%%%%%%%%%%%%%%%%%%%%%%%%%%%%%%%%%
%%%%%%%%%%%%%%%%%%%%%%%%%%%%%%%%%%%%%%%%%%%%%%%%%%%%%%%%%%%%%

\begin{figure}[htb]
\begin{center}
\includegraphics[width=60mm,height=50mm,angle=0,]{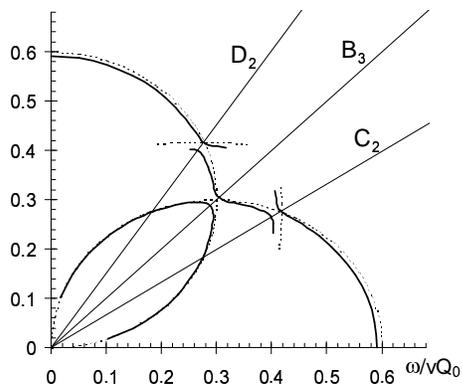}
\caption{Positions of the scattering peaks for $|K/Q_0|=0.6.$  Two
doublets at the BZ boundaries (points $C_{2}$ and $D_{2}$) in Fig.
\ref{BZNew}) are shifted from the high symmetry
directions.}\label{Kll_06}
\end{center}
\end{figure}

%\vspace{20mm}
Further increase of the arc radius $k\sin\varphi=-0.6Q_{0}$ (arc
$C_2B_3D_2$ in Fig.  \ref{BZNew}) leads to appearance of two points
$D_2$ and $C_2$ where the arc intersects with high symmetry lines (BZ
boundaries).  Each of these points generates an inter-band resonance
doublet, which coexists with the low frequency single peak
(Fig.\ref{Kll_06}).\\

%%%%%%%%%%%%%%%%%%%%%%%%%%%%%%%%%%%%%%%%%%%%%%%%%%%%%%%%%%%%%
\begin{figure}[htb]
\begin{center}
\includegraphics[width=60mm,height=50mm,angle=0,]{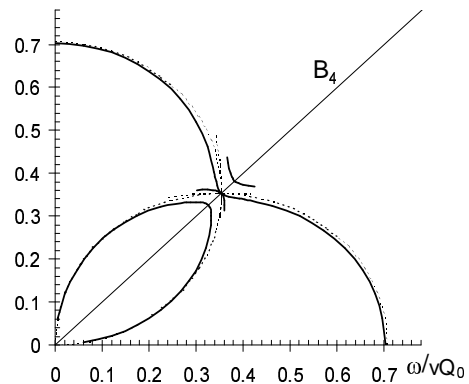}
\caption{Positions of the scattering peaks for
$|K/Q_0|=\sqrt{2}/2.$ Resonance triplet corresponding to point
$B_{4}$ in Fig. \ref{BZNew} (two of four frequencies remain
degenerate in our approximation). In IR experiments only one of
the triplet components is visible.}\label{Kll_07}
\end{center}
\end{figure}

%%%%%%%%%%%%%%%%%%%%%%%%%%%%%%%%%%%%%%%%%%%%%%%%%%%%%%%%%%%%%%
%\vspace{30mm}
In the case $k\sin\varphi=-\sqrt{2}Q_{0}/2$ the corresponding arc
includes the BZ corner $B_4.$ This is the point of a double
inter-array and inter-band resonance ({\bf v}).  Moreover, here the
two quartets described above coincide.  Therefore, there are three
lines in Fig.  \ref{Kll_07}.  The low-frequency line as its
high-frequency partner is two-fold degenerate while the central line
is four-fold degenerate.  We emphasize that each quartet manifests
itself in three lines, contrary to the IR absorption, where selection
rules make two of them invisible.\\

%%%%%%%%%%%%%%%%%%%%%%%%%%%%%%%%%%%%%%%%%%%%%%%%%%%%%%%%%%%%%%
\begin{figure}[htb]
\begin{center}
\includegraphics[width=60mm,height=50mm,angle=0,]{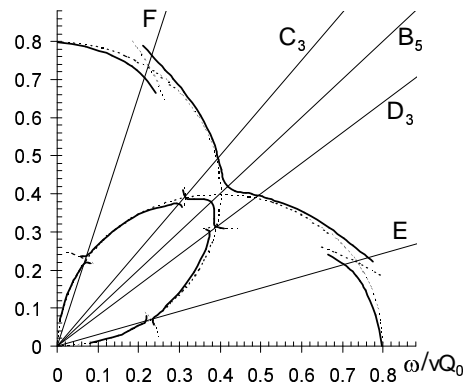}
\caption{Positions of the scattering peaks for $|K/Q_0|=0.8.$ Two
pairs of doublets appear corresponding to excitation of two pairs
of plasmons in the two arrays (points $E$ and $F$ in Fig.
\ref{BZNew}).}\label{Kll_08}
\end{center}
\end{figure}

%%%%%%%%%%%%%%%%%%%%%%%%%%%%%%%%%%%%%%%%%%%%%%%%%%%%%%%%%%%%%
The last case $k\sin\varphi=-0.8Q_{0}$ demonstrates one
more possibility related to inter-band resonance simultaneously in
two arrays ({\bf iii}).  Each point $E$ and $F$ generates (in the
corresponding direction) two doublets describing the inter-band
splitting
in different arrays (see Fig.  \ref{Kll_08}).\\

%\vspace{40mm}
%%%%%%%%%%%%%%%%%%%%%%%%%%%%%%%%%%%%%%%%%%%%%%%%%%%%%%%%%%%%%
%%%%%%%%%%%%%%%%%%%%%%%%%%%%%%%%%%%%%%%%%%%%%%%%%%%%%%%%%%%%%%
%%%%%%%%%%%%%%%%%%%%%%%%%%%%%%%%%%%%%%%%%%%%%%%%%%%%%%%%%%%%%%
\section{Conclusions}\label{sec:Conc}
%%%%%%%%%%%%%%%%%%%%%%%%%%%%%%%%%%%%%%%%%%%%%%%%%%%%%%
In conclusion, we studied inelastic UV Raman scattering on QCB. We
derived an effective Hamiltonian for QCB-light interaction which is
expressed via the same bose variables that the QCB itself.  With the
help of this Hamiltonian we calculated differential scattering cross
section as a function of detector orientation and scattered frequency. 
Scanning these parameters, one can observe a set of sharp peaks in the
scattering spectrum.  A number of peaks and their positions strongly
depend on the direction of the scattered wave vector.  This results in
a dimensional crossover.  It manifests itself in the splitting of the
peak frequencies and therefore in appearance of multiplets (mostly
doublets) instead of single lines in the scattering spectrum.\\

The sizes of peak splitting are determined by the nature of
interaction which lifts the corresponding degeneracy.  In the case of
initial inter-array degeneracy, the splitting is proportional to the
dimensionless interaction strength (\ref{small}).  The inter-band
splitting is proportional to the square of this interaction.  For a
chosen QCB parameters it is less than the inter-array interaction
strength in spite of an additional large multiplier $a/r_0.$ In all
cases the splitting increases with increasing the interaction in QCB
crosses.  The peak amplitudes are proportional to the Luttinger
parameter $g$ in a single nanotube.  Therefore strong
electron-electron interaction suppresses the peaks.\\

The effectiveness of UV scattering is related to the possibility
of changing continuously the excited plasmon frequency. Due to
other selection rules, some lines invisible in the IR absorption
spectrum, become observable in the UV scattering. UV scattering
spectroscopy enables one to restore parameters describing both
interaction in QCB crosses and electron-electron interaction in
a single QCB consituents.\\

Our studies of optical properties of QCB (this paper and
Refs.\onlinecite{KGKA3,K}) show that these nanoobjects possess unique
combination of optical spectra.  Firstly, they are active in IR and UV
frequency range.  Secondly, they may be observed in various kinds of
optical processes, namely direct and indirect absorption, diffraction,
energy loss transmission, and Raman-like spectroscopy.\\
%%%%%%%%%%%%%%%%%%%%%%%%%%%%%%%%%%%%%%%%%%%%%%%%%%%%%%%
\section*{Acknowledgements}
%%%%%%%%%%%%%%%%%%%%%%%%%%%%%%%%%%%%%%%%%%%%%%%%%%%%%%%
We thank Y. Imry for drawing our attention to the effectiveness of the
UV scattering in probing spectral properties of QCB.
%%%%%%%%%%%%%%%%%%%%%%%%%%%%%%%%%%%%%%%%%%%%%%%%%%%%%%%
\appendix
\section{QCB plasmons: kinematics of interference.}\label{sec:Kinem}
%%%%%%%%%%%%%%%%%%%%%%%%%%%%%%%%%%%%%%%%%%%%%%%%%%%%%%%%%
As it was mentioned in subsection \ref{sec:spectr}, standard description
(\ref{one-boson}), (\ref{freq}), of QCB plasmons in terms of array
plasmons fails at points ${\bf Q}$ lying on two specific groups of
lines in reciprocal space.  Nevertheless, QCB plasmons even at these
lines can be considered as  finite combinations of array plasmons
generated by the point ${\bf Q}$ by more complicated way.  Below we 
consider various types of the inrterference of array plasmons in QCB.\\

\begin{figure}[htb]
\begin{center}
\includegraphics[width=65mm,height=70mm,angle=0,]{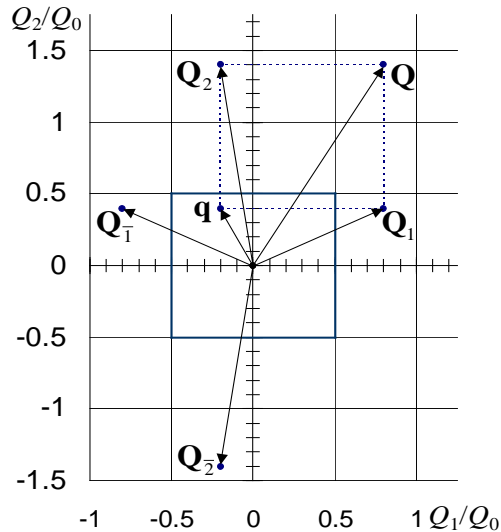}
\caption{QCB inverse space. The first BZ is bounded by a sooid
square}
\label{BZ-non}
\end{center}
\end{figure}

The first group of specific lines is formed by the high symmetry lines
$Q_j=p_jQ_0/2$ with $p_j$ an integer (lines parallel to coordinate axes
in Fig.  \ref{BZNew}).  The lines with $p_j=\pm 1$ include the BZ
boundaries.  In an ``empty lattice'' approximation, these lines are
degeneracy lines which separate $p_{j}$-th and $(p_{j}+1)$-th bands of
the $j$-th array.  Inter-array interaction mixes the array states,
lifts the (inter-band) degeneracy and splits corresponding
frequencies.  As a result, the QCB plasmons related to a point ${\bf
Q}$ lying ona high symmetry line (e.g. all $C,D$ points in Fig. 
\ref{BZNew}), are built from the array plasmons associated not only
with the points ${\bf Q}_{1,2}$ but also with the symmetric (with
respect to coordinate axes of a reciprocal space) points ${\bf
Q}_{\overline{1}}=(-Q_1,q_2)$ and ${\bf Q}_{\overline{2}}=(q_1,-Q_2)$
(see Fig.  \ref{BZ-non}).\\

Consider for definiteness the case $j=1$ and assume first that the
ratio $2Q_2/Q_0$ is non-integer.  In this case, the point ${\bf Q}$ 
generates three QCB plasmons.  The first one 
of them is a $|2,Q_2\rangle $ plasmon (see Eq.\ref{one-boson}) with
frequency $\omega_2.$ The two others
\begin{eqnarray}
    \label{states_1}
   |1,g,{\bf Q}_1\rangle &=&\frac{1}{\sqrt{2}}\left(
   |1,{\bf Q}_1\rangle +|1,{\bf Q}_{\overline{1}}\rangle
   \right),\nonumber\\
   |1,u,{\bf Q}_1\rangle&=&\frac{1}{\sqrt{2}}\left(
   |1,{\bf Q}_1\rangle-|1,{\bf Q}_{\overline{1}}\rangle
   \right)
\end{eqnarray}
are even or odd superpositions of the $1$-st array states from
the two first zones with eigenfrequencies
\begin{eqnarray}
    \label{freq_b_1}
    \omega_{1g}({\bf Q})&=&v|Q_1|,\nonumber\\
    \omega_{1u}({\bf Q})&=&\left(1-\frac{\phi^2a}{r_0}\right)v|Q_1|.
\end{eqnarray}
The case $j=2$ is described similarly after change $1\leftrightarrow2$
through all this paragraph.\\\\

%%%%%%%%%%%%%%%%%%%%%%%%%%%%%%%%%%%%%%%%%%%%%%%%%%%%%%%%%%%%%
\begin{figure}[htb]
\begin{center}
\includegraphics[width=75mm,height=65mm,angle=0,]{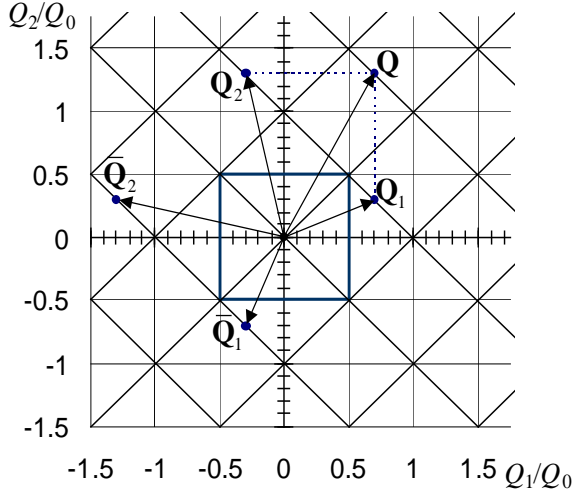}
\caption{Wave vectors involved in the resonance interaction. Resonant
lines are displayed by solid lines.}
\label{BZ-res}
\end{center}
\end{figure}

%%%%%%%%%%%%%%%%%%%%%%%%%%%%%%%%%%%%%%%%%%%%%%%%%%%%%%%%%%%%%%
The second set is formed by the resonant lines defined by the equation
$Q_{1}+ r Q_{2}=nQ_0,$ where $r=\pm 1$ and $n$ are two integer
parameters determining the line.  In Fig.  \ref{BZNew}, resonant lines
are square diagonals and lines parallel to them.  In an ``empty
lattice'' approximation, these lines are also degeneracy lines. 
However this degeneracy has more complicated nature.  Here, array
plasmons associated not only with the point ${\bf Q}$, but also array
plasmons associated with the dual point ${\bf
\overline{Q}}=(-rQ_2,-rQ_1)$ (see Fig.  \ref{BZ-res}) are involved in
the resonance.  Inter-array interaction mixes degenerate modes, lifts
the degeneracy, and splits degenerate frequencies.  As a result, the
QCB plasmons related to a point ${\bf Q}$ at the resonant line (e.g.
all $B,F,E$ points in Fig.  \ref{BZNew}), are built from the array
plasmons associated not only with the points ${\bf Q}_{1,2}$ but also
with the symmetric (with respect to one of the bisector lines of the
coordinate system of a reciprocal space) points ${\overline{\bf
Q}_1}=(-rQ_1,-rq_2)$ and ${\overline{\bf Q}_2}=(-rq_1,-rQ_2).$
Moreover, if the point ${\bf Q}$ lies at the intersection of two
resonant lines (like the point $B_{4}$ in Fig, \ref{BZNew}), the QCB
plasmons include also array states associated with the ${\overline{\bf
Q}}_{\overline j}$ points.\\

In the general case, the point ${\bf Q}$ lies only on one of the
resonant lines away from the high symmetry lines.  Here the point
${\bf Q}$ generates two pairs of even and odd superpositions of the
first and the second array states
\begin{eqnarray}
    \label{first_1}
   |g/u,{\bf Q}_1\rangle&=&\frac{1}{\sqrt{2}}\left(
   |1,{\bf Q}_1\rangle\pm |2,{\bf
   \overline{Q}}_1\rangle\right),\nonumber\\
   |g/u,{\bf Q}_2\rangle&=&\frac{1}{\sqrt{2}}\left(
   |2,{\bf Q}_2\rangle\pm |1,{\bf
   \overline{Q}}_2\rangle\right)
\end{eqnarray}
whose eigenfrequencies form two doublets
\begin{eqnarray}
\label{freq_a_1}
    \omega_{g/u}({\bf Q}_1)&=&\left(1\pm
    \frac{\phi}{2}\right)v|Q_1|,\nonumber\\
    \omega_{g/u}({\bf Q}_2)&=&\left(1\pm
    \frac{\phi}{2}\right)v|Q_2|.
    %\nonumber
\end{eqnarray}
\\
%%%%%%%%%%%%%%%%%%%%%%%%%%%%%%%%%%%%%%%%%%%%%%%%%%%%%%%%%%%%%
%%%%%%%%%%%%%%%%%%%%%%%%%%%%%%%%%%%%%%%%%%%%%%%%%%%%%%%%%%%%%%

The crossing points of resonant lines can also be divided into two
groups.  The first group is formed only by the crossing points of two
high symmetry lines (e.g. points $D_4,C_4$ in Fig.  \ref{BZNew}). 
Here each of the generated vectors ${\bf Q}_1$ and ${\bf Q}_2$
corresponds to a pair of array eigenstates belonging to two adjacent
$1D$ bands.  Thus an {\it inter-band} mixing is significant in both
two arrays.  The corresponding QCB plasmons are even or odd
combinations of $j$-th array plasmons ($j=1,2$)
\begin{eqnarray}
    \label{high_cross}
  |j,g,{\bf Q}_j\rangle &=& \frac{1}{2}\left(|j,{\bf Q}_{j}\rangle
  +|j,{\bf Q}_{\overline{j}}\rangle\right), \nonumber\\
  |j,u,{\bf Q}_j\rangle &=& \frac{1}{2}\left(|j,{\bf Q}_{j}\rangle-|j,{\bf
  Q}_{\overline{j}}\rangle\right),
\end{eqnarray}
with eigenfrequencies
\begin{eqnarray}
    \label{freq_b}
    \omega_{j,g}(Q_j)&=&v\frac{Q_0 p_j}{2},\nonumber\\
    \omega_{j,u}(Q_j)&=&\left(1-\frac{\phi^2a}{r_0}\right)v\frac{Q_0 p_j}{2}.
\end{eqnarray}
\\

The second group consists of the crosses of two resonant lines (point
$B_4$ in Fig.  \ref{BZNew}).  These point are always the crosses
of two high symmetry lines.  Here the QCB plasmons, generated by the
point ${\bf Q},$ form two quartets.  The first quartet is really
generated by the point ${\bf Q}_1.$ It consists of four symmetrized
combinations of single-array states
\begin{eqnarray}
  \label{quartet}
  |g,g,{\bf Q}_1\rangle =\frac{1}{2}&&\Big (
  |1,{\bf Q}_{1}\rangle+|1,{\bf Q}_{\overline{1}}\rangle+\nonumber\\
  &&|2,{\bf \overline{Q}}_{1,}\rangle+
  |2,{\bf \overline{Q}}_{\overline{1}}\rangle\Big );\nonumber\\
  |g,u,{\bf Q}_1\rangle =\frac{1}{2}&&\Big (
  |1,{\bf Q}_{1}\rangle+|1,{\bf Q}_{\overline{1}}\rangle-\nonumber\\
  &&|2,{\bf \overline{Q}}_{1}\rangle-
  |2,{\bf \overline{Q}}_{\overline{1}}\rangle\Big );\nonumber\\
  |u,g,{\bf Q}_1\rangle =\frac{1}{2}&&\Big (
  |1,{\bf Q}_{1}\rangle+|1,{\bf Q}_{\overline{1}}\rangle-\nonumber\\
  &&|2,{\bf \overline{Q}}_{1}\rangle-
  |2,{\bf \overline{Q}}_{\overline{1}}\rangle\Big );\nonumber\\
  |u,u,{\bf Q}_1\rangle =\frac{1}{2}&&\Big (
  |1,{\bf Q}_{1}\rangle-|1,{\bf Q}_{\overline{1}}\rangle-\nonumber\\
  &&|2,{\bf \overline{Q}}_{1}\rangle+|2,{\bf
  \overline{Q}}_{\overline{1}}\rangle\Big ).
\end{eqnarray}
QCB eigenstates of the second quartet are generated by the point ${\bf
Q}_2.$ They are obtained from equations (\ref{quartet}) by replacing
$1\leftrightarrow 2.$ Even array eigenstates are degenerate with
frequencies
\begin{equation}
    \label{evar}
    \omega_{g,g/u}({\bf Q}_{1,2})=\omega_{1,2},
\end{equation}
while the odd array eigenstates are split
\begin{equation}
    \label{odar}
    \omega_{u,g/u}({\bf Q}_{1,2})=(1\pm \phi)\omega_{1,2}.
\end{equation}
\\

Equations (\ref{states_1}) - (\ref{odar}) exhaust all cases of QCB
plasmons, generated by a point ${\bf Q}_2$ lying on a specific line,
via the interference of array plasmons.\\
%%%%%%%%%%%%%%%%%%%%%%%%%%%%%%%%%%%%%%%%%%%%%%%%%%%%%%%%%%%%%%%%%%%%%%%
\section{Derivation of the basic equations
of subsection \ref{sec:Light}.}\label{sec:Inter}
%%%%%%%%%%%%%%%%%%%%%%%%%%%%%%%%%%%%%%%%%%%%%%%%%%%%%%%
Light scattering on QCB is described by equations
(\ref{H-int-initial}) - (\ref{Pol_Matr1}).  In this Appendix we
briefly explain the main steps which lead to such a description.\\

\noindent \textbf{1.  Nanotube-light interaction, Eq.
(\ref{H-int-initial}).} We start with a consideration of a single
nanotube of the first array, interacting with an external
electromagnetic field.  The characteristic time of all nanotube
energies, including Coulomb interaction, is of order of an inverse
plasmon frequency.  The scattering process occurs during much shorter
time interval, which is of the order of an inverse photon frequency.
Hence, Coulomb interaction is irrelevant for the scattering processes.
This enables us to restrict ourselves to the kinetic part of the
nanotube Hamiltonian. Within the ${\bf k}-{\bf p}$ approximation,
this part is given by \cite{Ando3}
\begin{eqnarray}
h_{kin} = && v_F\int\frac{dx_{1} d\gamma}{2\pi}
          \times\nonumber\\
     &&\psi_{\alpha}^{\dag}(x_{1},r_{0},\gamma)
     \hbar{\bf{k}}\cdot
     {\boldsymbol\sigma}_{\alpha,\alpha'}
     \psi_{\alpha'}(x_{1},r_{0},\gamma).
       \label{H-kinetic-1}
\end{eqnarray}\\
\emph{In the presence of a magnetic field, one should add}
$e\textbf{A}/c$ \emph{to the electron momentum operator} $\hbar {\bf
k}.$ \textit{An additional part of the Hamiltonian}
(\ref{H-kinetic-1}) \emph{exactly coincides with the nanotube-light
interaction Hamiltonian}(\ref{H-int-initial}).\\
%%%%%%%%%%%%%%%%%%%%%%%%%%%%%%%%%%%%%%%%%%%%%%%%%%%%%%%%%%%%%%%%
\begin{figure}[htb]
\begin{center}
\includegraphics[width=65mm,height=60mm,angle=0,]{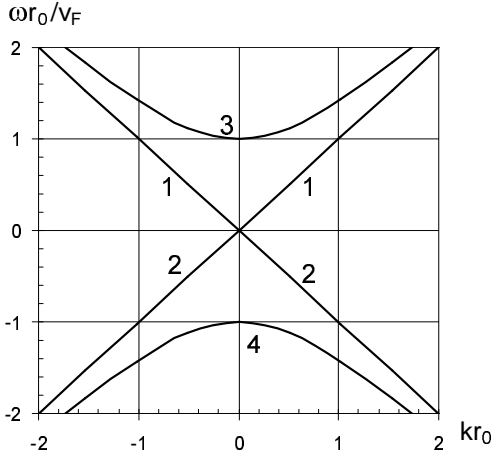}
 \caption{The low-energy part of the band structure of a metallic
 nanotube.  Lines $1,3$ ($2,4$) correspond to excitations with $p=+1$
 ($p=-1$).  Lines $1,2$ ($3,4$) correspond to excitations with orbital
 quantum number $m=0$ ($m=\pm 1$).}
  \label{CN-sp-2D}
\end{center}
\end{figure}
%%%%%%%%%%%%%%%%%%%%%%%%%%%%%%%%%%%%%%%%%%%%%%%%%%%%%%%%%%%%%%%%
Hamiltonian (\ref{H-kinetic-1}) is diagonalized by a two step
canonical transformation.  The first one is Fourier transformation
\begin{equation}
    \psi_{\alpha}(x_{1},\gamma)=
     \frac{1}{\sqrt{L}}
     \sum_{k,m}c_{\alpha,k,m}e^{ikx_{1}+im\gamma},
    \label{Fourier}
\end{equation}
where the orbital moment $m$ is restricted by the condition $|m|\leq
m_0=[\pi r_0/a_0]$ due to the finite number of honeycomb cells along
the nanotube perimeter ($a_{0}\sqrt{3}$ is the lattice constant).  The
second rotation is defined as
\begin{eqnarray*}
  c_{p,k,m}=\frac{1}{\sqrt{2}}
      \left(
       {c}_{akm}+p
       {e}^{i\phi_m}
       {c}_{bkm}
      \right),\ \ \  p=\pm,
  \label{c-AB--c-pm}
\end{eqnarray*}
where
$$
 \cos\phi_m=\frac{kr_0}{\sqrt{(kr_0)^2+m^2}},
 \ \ \ \ \
 \sin\phi_m=\frac{m}{\sqrt{(kr_0)^2+m^2}}.
$$
As a result, Hamiltonian takes the form
\begin{equation}\label{kin1}
H_{kin}=\sum_{p,k,m}\hbar \omega_p(k,m)c^\dag_{pkm}c_{pkm}
\end{equation}
with eigenfrequencies
\begin{equation}
    \omega_p(k,m)=pv_F \sqrt{k^2+\frac{m^2}{r_0^2}}.
    \label{eigenenergies}
\end{equation}\\

\noindent \textbf{2.  QCB-light interaction, Eq.
(\ref{Eff_Int_QCB}).} Substituting into Eq.  (\ref{H-int-initial}) the
scalar product ${\bf{A}}\cdot{\boldsymbol\sigma}$ in the form
\begin{eqnarray*}
 \left({\bf A}\cdot\boldsymbol\sigma\right)
 &=&
 \left(
      \begin{array}{cc}
      0     & A^{-}\\
      A^{+} & 0
      \end{array}
 \right),
 \label{sigma*A}
\end{eqnarray*}
where $A^{\pm}=A_1\pm iA_{\gamma},$ we can write the nanotube-light
interaction as
\begin{equation}
 H_{nl} = \frac{ev_F}{ c}\int\frac{dx_1d\gamma}{2\pi}
         \Big(
          \psi_A^{\dag}
          A^{-}
          \psi_B
          +h.c.
         \Big).
         \nonumber
 \label{H-int-pm}
\end{equation}
\\

We are interested in an effective QCB-light interaction Hamiltonian
obtained in second order of perturbation which describes transitions
between initial states $|i\rangle$ and final ones $|f\rangle.$ Initial
states are one-photon states of the electromagnetic field and the
electron ground state of the nanotube whereas final states consist of
one photon and an electron above the Fermi level.  It will be seen
later that just these states form a one-plasmon array states.  The
energy of incident photon $E=\hbar{ck}$ is much larger than the
excitation energies of the nanotube.  Therefore, absorption of the
incident photon and radiation of the scattered photon occur
practically without retardation.  All these result in the interaction
Hamiltonian
\begin{eqnarray}
  h_{int} &=&
       \frac{e^2}{\hbar c}
       \left(\frac{v_F}{c}\right)^2
       \int \frac{dx_1d\gamma}{2\pi}  \int
       \frac{dx'_1d\gamma'}{2\pi}
       \sum_{
       %|{i}\rangle,|{f}\rangle,
       |{v}\rangle}
       %\sum_{}
       \frac{1}{k}
       \times\nonumber\\&&
       %\Big[
       \Big(
       %|{f}\rangle\langle{f}|
       \psi^{\dag}_{a}(x'_1,\gamma')
       A^{-}(x'_1,0,0)
       |{v}\rangle
       \times\nonumber\\&&
       \langle{v}|
       \psi_b(x'_1,\gamma')
       \psi^{\dag}_b(x_1,\gamma)
       |{v}\rangle
       \times\nonumber\\&&
       \langle{v}|
       A^{+}(x_1,0,0)
       \psi_a(x_1,\gamma)
       %|{i}\rangle\langle{i}|
       +\nonumber\\&&{a}\leftrightarrow{b}\Big),
       %+h.c.
       %\Big]
  \label{H-eff-int-1}
\end{eqnarray}
which corresponds to the diagram shown in
Fig. \ref{Feinmann-Diagramm}a (there is no photons in an intermediate
state).\\

Consider now the matrix element $\langle{v}|
\psi_{\alpha}(x'_1,\gamma') \psi^{\dag}_{\alpha}(x_1,\gamma)
|{v}\rangle$ which enters this Hamiltonian. Due to our choice of
initial and final states, only diagonal elements with respect to
both virtual states $|v\rangle$ and sublattice indices $\alpha$
survive. In the $c_{\alpha,k,m}$ representation, (\ref{Fourier})
they have the form
\begin{eqnarray*}
 &&\langle{v}|
       \psi_{\alpha}(x'_1,\gamma')
       \psi^{\dag}_{\alpha}(x_1,\gamma)
 |{v}\rangle=\frac{1}{L}\sum_{k,k'}\sum_{m,m'}\times\nonumber\\&&
    \langle{v}|
     c_{\alpha,k,m}
     c^{\dag}_{\alpha,k',m'}
    |{v}\rangle
    e^{ikx_1-ik'x'_1+im\gamma-im'\gamma'}.
 \label{n-alpha}
\end{eqnarray*}
Internal matrix elements on the r.h.s. of the latter equation are
\begin{equation}
    \langle{v}|
     c_{\alpha,k,m}
     c^{\dag}_{\alpha,k',m'}
    |{v}\rangle
    =\frac{1}{2}\delta_{k,m}\delta_{k',m'}
    \nonumber
    \label{internal}
\end{equation}
(here the symmetry property $n\left(E_{-}(k,m)\right)+
n\left(E_{+}(k,m)\right)=1$ is used). Therefore
\begin{equation}
    \langle{v}|
       \psi_{\alpha}(x'_1,\gamma')
       \psi^{\dag}_{\alpha}(x_1,\gamma)
     |{v}\rangle=\pi
    \delta(x_1-x'_1)
    S(\gamma-\gamma'),
    \nonumber
    \label{me}
\end{equation}
where
\begin{equation}
    S(\gamma)=\frac{1}{2\pi}\sum_{m=-m_0}^{m_0}e^{im\gamma}.
    \nonumber
    \label{gamma}
\end{equation}
\\

Thus, the interaction (\ref{H-eff-int-1})  takes the form
\begin{eqnarray*}
       h_{int} =
       \frac{e^2}{\hbar c}
       \left(\frac{v_F}{c}\right)^2
       \int\frac{dx_1d\gamma d\gamma'}{4\pi k}
       S(\gamma-\gamma')
       \times\nonumber\\
       %\Big(
       %|{f}\rangle\langle{f}|
       A^{-}(x_1,0,0)
       A^{+}(x_1,0,0)\sum_{\alpha}
       \psi^{\dag}_{\alpha}(x_1,\gamma')
       \psi_{\alpha}(x_1,\gamma).
       %|{i}\rangle\langle{i}|
       %+\nonumber\\
       %{A}\leftrightarrow{B}
       %+h.c.
       %\Big).
  \label{H-eff-int-2}
\end{eqnarray*}
The field dependent factor here is
\begin{eqnarray}
       A^{-}(x_1,0,0)A^{+}(x_1,0,0)=
       A_1^2(x_1,0,0)+\nonumber\\
       A^2(x_1,0,0)\sin(\gamma_{A}-\gamma')
       \sin(\gamma_A-\gamma),
    \label{field}
\end{eqnarray}
and $(A_{1},A,\gamma_{A})$ are cylindrical components of the vector
potential ${\bf A}$ (\ref{VecPot}).  Taking into account the angular
dependence of the field (\ref{field}), we can omit it in the electron
operators.  Indeed, according to Eq.  (\ref{eigenenergies}) (see also
Fig.  \ref{CN-sp-2D}) in an energy-momentum region where we work, the
$m=1$ spectral band with nonzero orbital moment is separated from that
with $m=0$ by an energy of order $\hbar v_{F}/r_{0},$ which
is much larger than the QCB plasmon energy.  Keeping only the zero
moment field operators which form the electron density operator
\begin{equation}
    \label{ro}
    \sum_{\alpha}\psi^{\dag}_{\alpha}(x_1)\psi_{\alpha}(x_1)=
    \rho(x_1)\equiv \sqrt{2}\partial_{x_1}\theta(x_1).
    \nonumber
\end{equation}
and integrating over $\gamma,\gamma',$ we obtain an interaction
Hamiltonian in the form
\begin{equation}
 h_{int} =\frac{\sqrt{2}}{4k}
 \frac{e^2}{\hbar c}
 \left(\frac{v_F}{c}\right)^2
 \int dx_1
 \partial_{x_1}\theta(x_1)
 {\bf{A}}_1^2(x_1,0,0),
 \label{Effect-int-3}
\end{equation}
where
\begin{equation}
 {\bf{A}}_1={\bf{A}}+
 (\sqrt{2}-1)A_1{\bf{e}}_1.
 \nonumber
 \label{A-1}
\end{equation}
\emph{Straightforward generalization of this expression to the QCB case
leads exactly to the Hamiltonian} (\ref{Eff_Int_QCB}).\\

\noindent \textbf{3.  Polarization matrix, Eq.  (\ref{Pol_Matr1}).} To
study the scattering process, we should modify the last expression for
the interaction Hamiltonian.  To proceed further, we define Fourier
transforms $\theta_{j, {\bf Q}_j}$ of the bosonic fields
\begin{eqnarray}
  \theta_1(x_1,n_2a) &=&
  \frac{1}{\sqrt{{N}{L}}}
  \sum_{{\bf Q}_1}
  \theta_{1,-{\bf Q}_1}
  \times\nonumber\\&&\times
  e^{-iQ_1x_1-iq_2n_2a},
  \nonumber\\
  \theta_2(n_1a,x_2) &=&
  \frac{1}{\sqrt{{N}{L}}}
  \sum_{{\bf Q}_2}
  \theta_{2,-{\bf Q_2}}
  \times\nonumber\\&&\times
  e^{-iq_1n_1a-iQ_2x_2}.
  \label{theta-Fourier}
\end{eqnarray}
Here $N=L/a$ is the number of QCB cells in both directions.  The
electromagnetic field also can be expanded in a sum of harmonics with
a wave vector ${\bf k}$ and polarization $\lambda=||,\bot,$
\begin{eqnarray}
 {\bf{A}}({\bf{r}}) &=&
 \sum_{{\bf{k}}\lambda}
 {\bf{n}}_{{\bf{k}}\lambda}A_{{\bf{k}}\lambda}
 e^{i{\bf{kr}}}.
 \label{A-Fourier}
\end{eqnarray}
The polarization vectors
\begin{eqnarray*}
{\bf{n}}_{{\bf{k}}||} &=&
 \frac{ik_2{\bf{e}}_1}{\sqrt{k_1^2+k_2^2}}-
 \frac{ik_1{\bf{e}}_2}{\sqrt{k_1^2+k_2^2}},
 \\
{\bf{n}}_{{\bf{k}}\bot} &=&
 \frac{k_1k_3{\bf{e}}_1}{k\sqrt{k_1^2+k_2^2}}+
 \frac{k_2k_3{\bf{e}}_1}{k\sqrt{k_1^2+k_2^2}}-
 \frac{\sqrt{k_1^2+k_2^2}}{k}{\bf{e}}_3,
\end{eqnarray*}
are normalized, $\left|{\bf{n}}_{{\bf{k}}\lambda}\right|=1,$
and satisfy the orthogonality conditions,
${\bf{n}}_{{\bf{k}}\lambda}\cdot{\bf{k}}=
 {\bf{n}}_{{\bf{k}}||}\cdot{\bf{n}}_{{\bf{k}}\bot}=0.$ The
field operators $A_{{\bf{k}}\lambda}$ satisfy the condition
$A^{\dag}_{{\bf{k}}\lambda}=A_{-{\bf{k}}\lambda}$, so that
${\bf{A}}^{\dag}({\bf{r}})={\bf{A}}({\bf{r}})$.\\

Substituting equations (\ref{theta-Fourier}), and
(\ref{A-Fourier}) into the Hamiltonian (\ref{Eff_Int_QCB}), we
obtain
\begin{eqnarray}
&&H_{int}=
 -i\frac{\sqrt{2NL}}{4k}
 \frac{e^2}{\hbar c}
 \left(\frac{v_F}{c}\right)^2
 \sum_{{\bf{k}},{\bf{k}'},{\bf{Q}}}\sum_{j,\lambda,\lambda'}
 \times
 \nonumber\\
 &&P_{j;\lambda',\lambda}\left(\frac{{\bf k}'}{k'},\frac{{\bf
    k}}{k}\right)Q_j
 A^{\dag}_{{\bf k}',\lambda'}
 A_{{\bf{k}},\lambda}
  \theta_{j,-{\bf Q}_j}.
 \label{Eff-Int-k}
\end{eqnarray}
Here
\begin{equation}
    \label{polariz}
    P_{j;\lambda',\lambda}\left(\frac{{\bf k}'}{k'},\frac{{\bf
    k}}{k}\right)=({\boldsymbol{\kappa}}^*_{j,{\bf k}',\lambda'}
  \cdot
  {{\boldsymbol{\kappa}}}_{j,{\bf k},\lambda})
\end{equation}
is the polarization matrix, ${\bf Q}={\bf q}+{\bf m},$ and
\begin{eqnarray*}
 {{\boldsymbol{\kappa}}}_{j,{\bf{k}},\lambda}
  &=&
  {\bf{n}}_{{\bf{k}}\lambda}+
  \left(\sqrt{2}-1\right)
  \left({\bf{n}}_{{\bf{k}}\lambda}\cdot{\bf{e}}_j\right){\bf{e}}_j.
 \label{ej}
\end{eqnarray*}
\emph{In the case of normal incidence,} Eqs. (\ref{polariz}),
(\ref{ej})
\emph{result in the form} (\ref{Pol_Matr1}) \emph{of polarization matrix.}\\

\noindent \textbf{4.  Scattering Hamiltonian, Eq.  (\ref{Fin_Int}).}
As a next step, we express the Fourier transforms of the Bose fields
$\theta$ via creation ($a^{\dag}$) and annihilation ($a$) operators of
the array plasmons
\begin{equation}
  \theta_{j,-{\bf Q}_j} =
  \sqrt{\frac{g}{2|Q_j|}}
  \left(
       a_{j,-{\bf Q}_j}+
       a^{\dag}_{j,{\bf Q}_j}
  \right).
  \nonumber
  \label{a-opers}
\end{equation}
The electromagnetic field amplitudes $A_{{\bf k},\lambda}$ should
also be expressed via photon creation ($c^{\dag}$) and
annihilation ($c$) operators
\begin{eqnarray*}
 A_{{\bf{k}},\lambda}(t) &=&\sqrt{
      \frac{\hbar{c}}{2Vk}}
      \left(
    c_{{\bf{k}},\lambda}
    +
    c_{-{\bf{k}},\lambda}^{\dag}
      \right).
      \label{A-opers}
\end{eqnarray*}
Substituting these expansions into Eq.  (\ref{Eff-Int-k}) we obtain
the final form of the effective interaction.
\emph{In the case of normal incidence} (${\bf{K}}={\bf{0}}$),
\emph{this interaction is written as}
\begin{eqnarray}
 H_{int} =&&
 \frac{i\sqrt{gNL}}{V}
 \left(\frac{ev_F}{2ck}\right)^2\times\nonumber\\
 &&\sum_{{\bf{k}},{\bf{k}'}}\sum_{j,\lambda,\lambda'}
 h_{\lambda',\lambda}({\bf K}')
 c^{\dag}_{{\bf{k'}},{\lambda}'}
 c_{{\bf{k}},\lambda},
 \label{Eff-Int-ck-1}
\end{eqnarray}
\emph{where} $h_{\lambda',\lambda}({\bf K}')$ \emph{is the
Hamiltonian} (\ref{Fin_Int}), \emph{where}
$\lambda_f,\lambda_i,{\bf K}$ \emph{are replaced
by} $\lambda',\lambda,{\bf K}'.$\\

\noindent \textbf{5.  Scattering cross section, Eqs. 
(\ref{eq:cross1}) - (\ref{Fin_Cross}).} Standard procedure based on
the Fermi golden rule leads to the following expression of the
differential scattering cross section per unit QCB square
\begin{equation}
 \frac{1}{L^2}\frac{d\sigma}{d\omega do}=\frac{1}{\pi}
 \left(\frac{Vk_f}{Lc\hbar} \right)^2
    \overline{\big|\big\langle f\big|H_{int}\big|i\big \rangle\big|^2}
      \delta\left(\frac{\varepsilon_i-\varepsilon_f}{\hbar}\right)
 \label{cross-def}
\end{equation}
(here bar denotes averaging with respect to polarization of both
incident light and scattered quanta).  Choose an initial ket-state
$\big | i \big \rangle$ such that it contains an incident photon with
momentum ${\bf k}_i,$ frequency $\Omega_i=ck_i,$ and polarization
$\lambda_i,$ and does not contain any QCB plasmon.  This state can be
written as $\big | i \big \rangle = | {\bf k}_i\rangle_l \bigotimes| 0
\rangle_p,$ where $| 0 \rangle_p$ is the plasmon vacuum, $| {\bf
k}_i\rangle_l=c^{\dag}_{{\bf k}_i,\lambda_i}|0\rangle_l,$ and
$|0\rangle_l$ is the photon vacuum.  A final bra-state $ \big \langle
f \big |$ contains a scattered photon with momentum ${\bf k}_f,$
frequency $\Omega_f=ck_f,$ and polarization $\lambda_f.$ It contains
also a QCB plasmon $P$ with the frequency $\omega_P$ (its quantum numbers
will be specified below).  The final state is written as $ \big
\langle f \big | = \langle P |\bigotimes \langle {\bf k}_f |,$ where
$\langle{\bf k}_f |=\left(c^{\dag}_{{\bf
k}_f,\lambda_f}|0\rangle_l\right)^{\dag}.$ \\

A matrix element of the interaction which enters Eq.
(\ref{cross-def}), is
\begin{equation}
    \label{MatrEl}
    \big\langle f\big|H_{int}\big|i\big \rangle=
    \langle P|\overline{H}_{int}|0\rangle_p,
\end{equation}
where
\begin{eqnarray}
    \overline{H}_{int}\equiv\left\langle{\bf k}_f,\lambda_f\left|
    H_{int}
    \right|{\bf k}_i,\lambda_i\right\rangle=&&\nonumber\\
    -\frac{i\sqrt{gNL}}{V}
    \left(\frac{ev_F}{2ck}\right)^2
    \sum_{j,{\bf Q}}
    \frac{Q_j}{\sqrt{|Q_j|}}
    \delta_{{\bf K}_f,{\bf K}_i-{\bf Q}}
    \times&&\nonumber\\
    P_{j;\lambda_f,\lambda_i}\left(\frac{{\bf k}_f}{k_f},\frac{{\bf
    k}_i}{k_i}\right)\left(
    a_{j,-{\bf Q}_j}+
    a^{\dag}_{j,{\bf Q}_j}
    \right).&&
    \label{matel}
\end{eqnarray}
\emph{In the case of normal incidence} Eqs. (\ref{cross-def}) -
(\ref{matel}) \emph{are equivalent to} Eqs. (\ref{Eff_Int_QCB}) -
(\ref{Fin_Int}).
%%%%%%%%%%%%%%%%%%%%%%%%%%%%%%%%%%%%%%%%%%%%%%%%%%%%%%%

%%%%%%%%%%%%%%%%%%%%%%%%%%%%%%%%%%%%%%%
\end{document}